\documentclass[reprint,amsmath,amssymb,aps,prl,superscriptaddress]{revtex4-2}
\pdfoutput=1

\usepackage{graphicx}%
\usepackage{hyperref}
\usepackage{cleveref}
\crefname{figure}{Fig.}{Figs.}
\Crefname{figure}{Fig.}{Figs.}
\crefname{section}{Sec.}{Secs.}
\Crefname{section}{Sec.}{Secs.}
\newcommand{\smref}[1]{see~\cref{#1}}
\usepackage{placeins}

\begin{document}

\title{Sub-MHz SWAP spectroscopy of two-level systems in superconducting qubits}

\author{Joaquin Minguzzi}
\affiliation{Amazon Center for Quantum Computing, Pasadena, CA 91106, USA}
\author{John Clai Owens}
\affiliation{Amazon Center for Quantum Computing, Pasadena, CA 91106, USA}
\author{Oskar Painter}
\affiliation{Amazon Center for Quantum Computing, Pasadena, CA 91106, USA}
\affiliation{Institute for Quantum Information and Matter, California Institute of Technology, Pasadena, CA 91125, USA}
\affiliation{Thomas J. Watson, Sr., Laboratory of Applied Physics, California Institute of Technology, Pasadena, CA 91125, USA}
\author{Matthew H. Matheny}
\affiliation{Amazon Center for Quantum Computing, Pasadena, CA 91106, USA}
\author{Loren J. Swenson}
\affiliation{Amazon Center for Quantum Computing, Pasadena, CA 91106, USA}

\begin{abstract}
An outstanding challenge in superconducting quantum circuits is mitigating loss from two-level system (TLS) defects.
SWAP spectroscopy with flux-tunable qubits can detect individual TLSs, but the Rabi chevrons that signal coherent exchange have been observed only for the most strongly coupled defects.
Here, we introduce a sequence with fast flux pulses only at state preparation and readout, switched off during the energy-relaxation delay for finer frequency resolution and longer interaction times.
We detect Rabi chevrons at swap rates as low as $0.1$~MHz, an order of magnitude lower than previously reported.
In the dressed-state picture, we fit the coherent swap oscillations to extract the TLS energy-relaxation time.
These Rabi chevrons reveal coherent TLSs with a median energy-relaxation time of $10\ \mu$s, at coupling strengths previously associated with incoherent TLSs.
The swap coherence times exceed the qubit $T_2^*$ limited by flux noise, indicating that the dressed gap protects them from low-frequency $1/f$ noise.
\end{abstract}

\maketitle

\begin{figure}[!t]
\centering
\includegraphics[width=0.95\columnwidth]{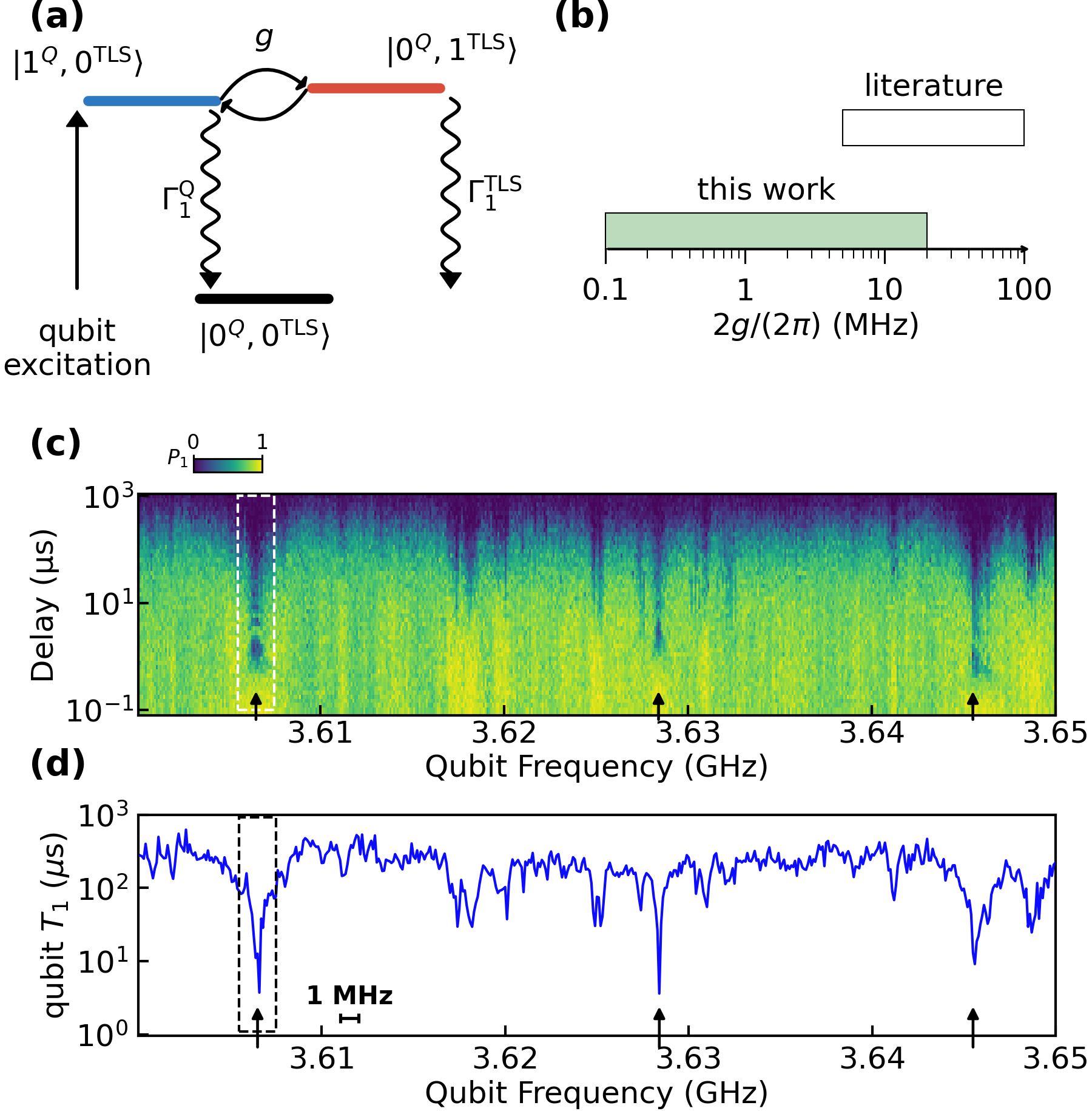}
\caption{\protect\input{figure1_caption.tex}}
\label{fig:figure1}
\end{figure}

As the coherence times of superconducting qubit devices improve~\cite{Siddiqi2021}, the signatures of two-level system (TLS) defects become subtler and demand more sensitive detection methods~\cite{Martinis2005,Muller2019}.
SWAP spectroscopy, also called TLS spectroscopy, probes individual TLSs with a flux-tunable qubit that is excited and then tuned across frequency to detect resonant energy exchange~\cite{Simmonds2004,Cooper2004}.
This method is now standard for assessing device quality, from multiqubit processors~\cite{Klimov2018,Zanuz2025} to wafer-scale fabrication~\cite{VanDamme2024,Ke2025}, yielding defect coherence times and coupling strengths $g$~\cite{Osman2023,Weeden2025,Lisenfeld2019,Lisenfeld2025}.
Because $g$ scales with the local electric field, TLSs inside the Josephson junction (JJ) barrier couple strongly, with $g$ exceeding the energy-relaxation rates of the TLS ($\Gamma_1^{\rm TLS}$) and of the qubit ($\Gamma_1^{\rm Q}$); see~\cref{fig:figure1}.
In this regime the interaction is coherent, producing Rabi exchange between qubit and TLS and the distinctive chevron pattern of SWAP spectroscopy (see~\cref{fig:figure1}(c)).
Recent studies of coherent TLSs, from phonon-engineered bandgap structures~\cite{Chen2024} to merged-element transmon devices~\cite{Mamin2021,Daum2025}, have also remained in this strongly coupled regime.

Beyond these strongly coupled TLSs, more coherent qubits uncover a regime where the coupling is much weaker~\cite{Barends2013,Klimov2018}.
Yet to date, coherent detection of TLSs, i.e., resolving Rabi exchange, has been limited to interaction times below $1\ \mu$s and swap rates $2g/(2\pi)$ above several MHz~\cite{Shalibo2010,Chen2024} (see~\cref{fig:figure1}(b)).
In this work, we resolve coherent qubit-TLS swap oscillations at rates $2g/(2\pi)$ as low as $0.1$~MHz in the regime previously associated with weakly coupled TLSs.
We use a fast-flux-assisted (FFA) pulse sequence to enable SWAP spectroscopy with a $0.1$~MHz frequency step and delay times up to $1$~ms.
Across 16 flux-tunable transmon qubits with an I-mon design on 5 aluminum-on-silicon chips, we characterized 118 coherent TLSs whose relaxation dynamics are described by the dressed-state picture~\cite{Muller2009}.
The measured swap coherence times exceed the qubit $T_2^*$, consistent with the dressed gap suppressing the linear coupling to low-frequency noise~\cite{Makhlin2003,Muller2009}.

To illustrate the sensitivity of our method, we focus on the sub-MHz Rabi chevrons in the heatmap of~\cref{fig:figure1}(c), where the qubit excited-state population $P_1$ (color scale) is plotted versus qubit frequency and delay.
Within a $50$~MHz window, three Rabi chevrons with coherent oscillations beyond $10\ \mu$s are marked by black arrows.
In the $T_1$ spectrum of~\cref{fig:figure1}(d), obtained from exponential-decay fits to each frequency slice, the same TLS resonances appear as localized dropouts with fitted $T_1$ values as low as a few microseconds.
Previously, similar dropouts were attributed to weakly coupled TLSs at the superconducting device interfaces~\cite{Lisenfeld2019,Osman2023,Weeden2025}, where $g$ falls below $\Gamma_1^{\rm TLS}$ and the interaction becomes incoherent.
Here, at the TLS resonances, the $P_1$ decay is nonexponential, and the exponential fit captures only the initial decay on a timescale of $\sim 1/(2g)$, a few microseconds for sub-MHz coupling strengths.
Away from the TLS resonances, the decay is purely exponential with a baseline $T_1$ of several hundred microseconds.

\begin{figure}[t]
\centering
\includegraphics[width=0.95\columnwidth]{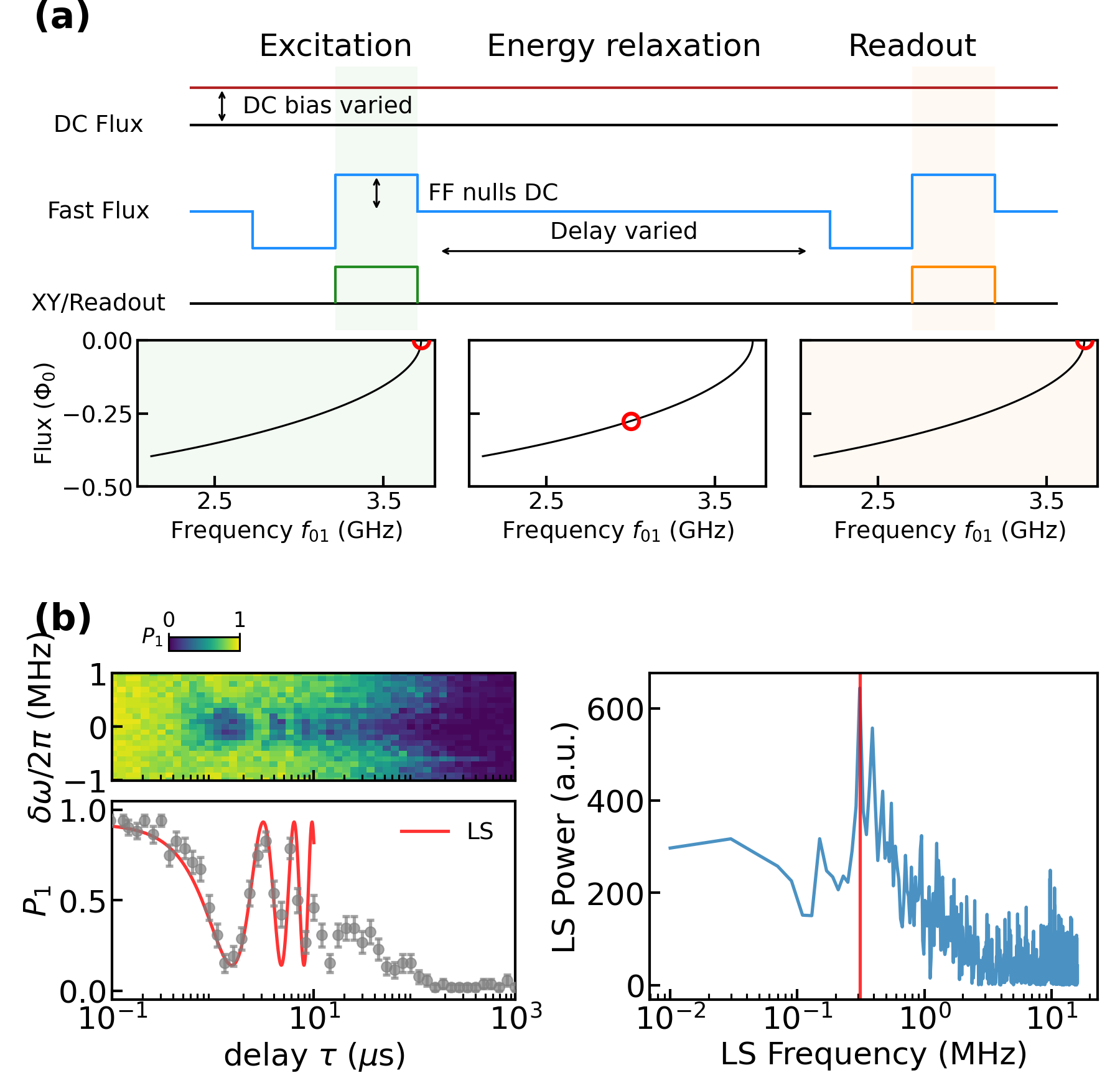}
\caption{\protect\input{figure2_caption.tex}}
\label{fig:figure2}
\end{figure}

Characterizing these TLS defects requires resolving coherent dynamics with frequency precision below $1$~MHz over delay times from microseconds to milliseconds.
In conventional SWAP spectroscopy, DC flux biases the qubit at the flux-insensitive sweet spot for state preparation and readout, while a fast-flux (FF) pulse brings the qubit to the spectroscopy frequency and holds it there for variable delays.
However, long flux pulses are subject to distortions from the bias tee, filtering elements, and impedance mismatches in the flux control line\footnote{DC and FF lines are combined through a cryogenic bias tee to control the magnetic flux through the transmon SQUID loop, with the capacitor removed from the FF side, similar to the experiments in Ref.~\cite{Levine2024}.}.
State-of-the-art methods calibrate these distortions on timescales up to $\sim 10\ \mu$s, achieving residual frequency errors down to $\sim 1$~MHz~\cite{Hellings2025}.
Both the calibration timescale and the residual frequency error are set by the qubit $T_2^*$-limited sensitivity of \textit{in situ} Ramsey and spectroscopy experiments used to characterize the flux line.

Our higher detection sensitivity comes from the FFA-SWAP pulse sequence shown in~\cref{fig:figure2}(a).
Here, DC flux sets the spectroscopy frequency, and two short FF pulses bring the qubit to its flux sweet spot for state preparation and readout.
By switching the FF completely off during the $T_1$ delay, we eliminate the need to calibrate long flux pulses and reduce flux noise to the level set by the DC source alone.
The short FF pulses are calibrated with IIR and FIR predistortion filters~\cite{Rol2020} and are shaped to be net zero, carrying reduced spectral weight at $\sim$kHz frequencies and thereby suppressing slow transients in the flux control line.
Residual pulse imperfections in principle only reduce the overall contrast of $P_1$, but pulse transients can also distort the data at short and long delays, affecting both the shape of the Rabi chevron and the exponential decay away from it.
Each pulse dissipates power in the line components, heating them and changing their impedance; the resulting cooling transients, of $100\ \mu$s or longer~\cite{Paluch2024}, can affect subsequent FF pulses within and between sequences.
We therefore keep the FF pulses as short as possible.

Beyond pulse design, a second key ingredient for resolving TLS parameters is time-domain sampling.
Other approaches estimate $T_1$ from a single delay time, trading accuracy for speed; sampling at multiple delays preserves accuracy and provides the time-domain information needed to detect coherent oscillations.
We use 51 logarithmically spaced delay times $\tau$ spanning $10^{-1}$--$10^3\ \mu$s to capture the full dynamics of $P_1(\tau)$ and avoid misinterpreting oscillatory decay as purely incoherent.
The Lomb--Scargle (LS) periodogram is well suited to this sampling, handling both uneven time spacing and measurement uncertainties (per-point error bars) through a weighted least-squares fit~\cite{VanderPlas2018}.
The frequency at which the LS power peaks serves as an initial guess for the oscillatory-decay model used to fit the $P_1$ dynamics; see~\cref{fig:figure2}(b).

\begin{figure}[t]
\centering
\includegraphics[width=0.95\columnwidth]{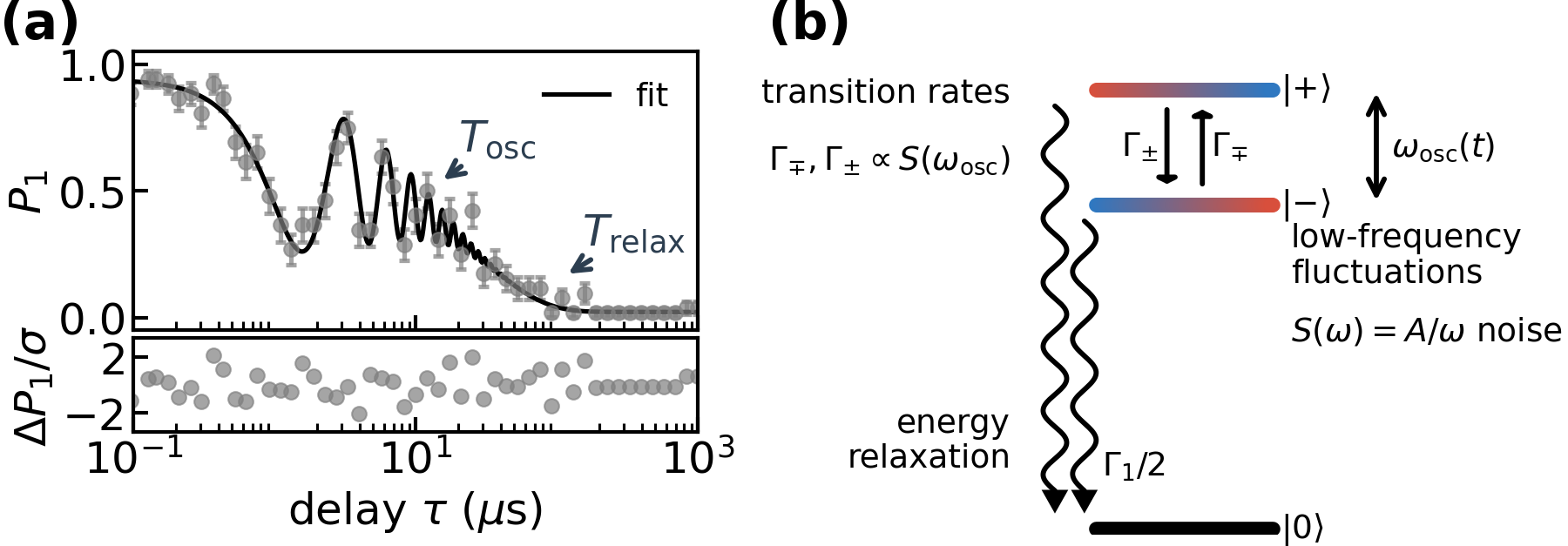}
\caption{\protect\input{figure3_caption}}
\label{fig:figure3}
\end{figure}

The $P_1$ time trace at the TLS resonance reveals two distinct decay components, both captured by a single fit function, as shown in~\cref{fig:figure3}(a).
The first is a fast oscillation, and the second a slow, purely exponential tail that persists after the oscillation has damped out.
We fit the trace to the model function $P_1(\tau)=c+a_{\rm osc}\sin(\omega_{\rm osc}\tau+\phi)e^{-\tau/T_{\rm osc}}+a_{\rm relax}e^{-\tau/T_{\rm relax}}$, with offset $c$, angular frequency $\omega_{\rm osc}$, phase $\phi$, amplitudes $a_{\rm osc}$, $a_{\rm relax}$, and times $T_{\rm osc}$, $T_{\rm relax}$~\cite{Muller2009,Burnett2019} (\smref{sec:supp_fitting}).
The residuals in the lower strip of~\cref{fig:figure3}(a) lie within $\pm2\sigma$ of the fit, where $\sigma$ is the per-point error bar, and give a reduced chi-squared $\chi^2/\nu=1.00$ for this seven-parameter model ($\nu=44$ degrees of freedom).
Among the fitted parameters are an oscillation frequency $\omega_{\rm osc}/2\pi=0.32\pm0.01$~MHz, a swap coherence time $T_{\rm osc}=9\pm3\ \mu$s, and a relaxation time $T_{\rm relax}=30\pm3\ \mu$s, where the uncertainties are standard errors from the least-squares fit.
Here $T_{\rm relax}$ is much longer than $T_{\rm osc}$, a separation of timescales we can resolve because both the qubit and TLS energy-relaxation times are long.
This illustrates a qubit-TLS Rabi exchange with a coherent swap an order of magnitude slower and an order of magnitude longer lived than in previous experiments with strongly coupled TLSs~\cite{Shalibo2010,Chen2024}.

The dressed-state picture~\cite{Muller2009}, sketched in~\cref{fig:figure3}(b), maps the fitted parameters to physical quantities.
The qubit and TLS hybridize into dressed states $\vert\pm\rangle$ separated by $\omega_{\rm osc}=\sqrt{(2g)^2+\delta\omega^2}$, where $\delta\omega/2\pi=f_{01}-f_{\rm TLS}$ is the detuning.
The energy-relaxation rates of the qubit and TLS, $\Gamma_1^{\rm Q}$ and $\Gamma_1^{\rm TLS}$ (see~\cref{fig:figure1}(a)), add to a total rate $\Gamma_1=\Gamma_1^{\rm Q}+\Gamma_1^{\rm TLS}$.
At resonance ($\delta\omega=0$) the gap is $\omega_{\rm osc}=2g$ and each dressed state, an equal superposition of qubit and TLS excitation, relaxes to $\vert0\rangle$ at $\Gamma_1/2$: this is the rate of the exponential tail, $\Gamma_{\rm relax}\equiv1/T_{\rm relax}=\Gamma_1/2$.
The oscillation decays faster, at $\Gamma_{\rm osc}\equiv1/T_{\rm osc}=\Gamma_1/2+\Gamma_v$, where $\Gamma_v$ quantifies the excess decoherence of the dressed gap beyond energy relaxation.
Two channels contribute to $\Gamma_v$: Markovian transitions between the dressed states, at rates $\Gamma_\pm$ and $\Gamma_\mp$ driven by noise at the gap frequency $S(\omega_{\rm osc})$, and non-Markovian dephasing from low-frequency $1/f$ noise, which the gap suppresses to quadratic order in the detuning fluctuations, $\omega_{\rm osc}(t)\approx2g+[\delta\omega(t)]^2/(4g)$ (see~\cref{sec:supp_dephasing}).
The resonant swap dynamics therefore yield the coupling $2g$, the total relaxation rate $\Gamma_1$, and the dressed-gap decoherence rate $\Gamma_v$.

\begin{figure}[!t]
\centering
\includegraphics[width=0.95\columnwidth]{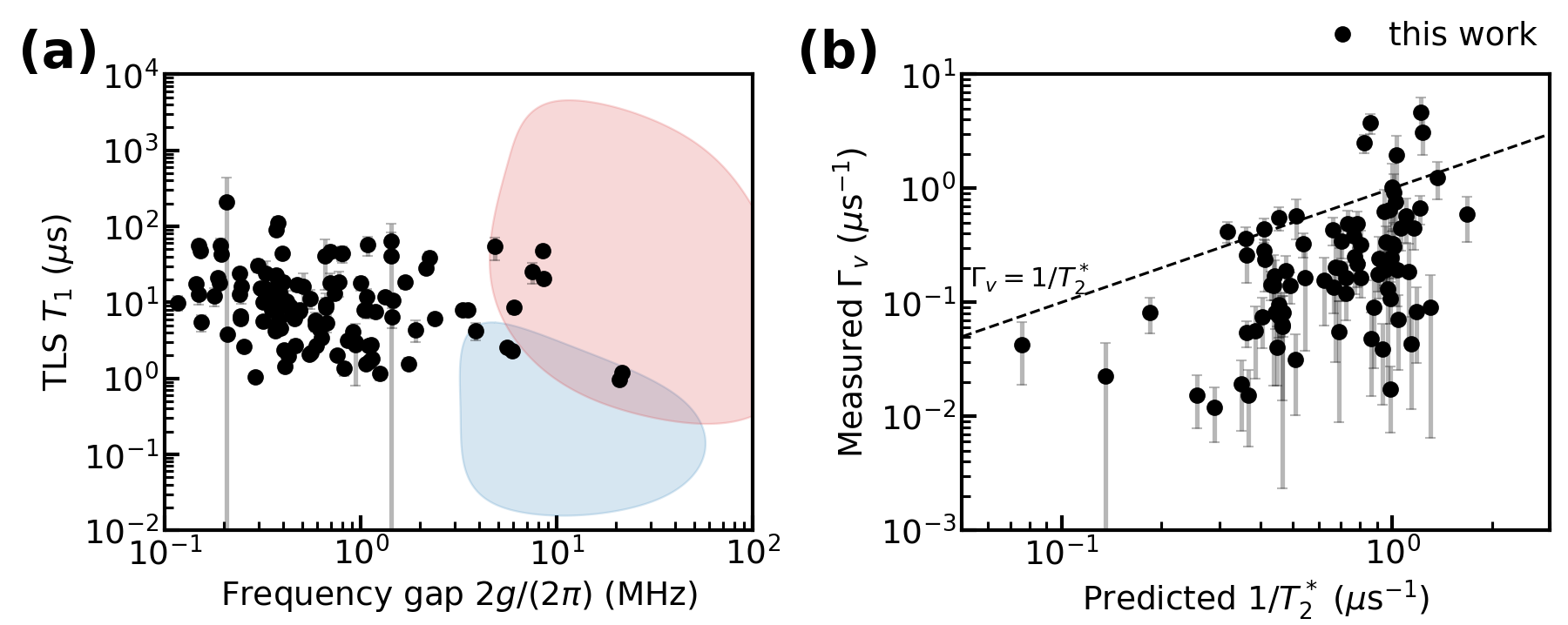}
\caption{\protect\input{figure4_caption}}
\label{fig:figure4}
\end{figure}

For each Rabi chevron we extracted the dressed gap $2g$ and the TLS relaxation time $T_1^{\rm TLS}$, combining the fitted $T_{\rm relax}$ with the qubit relaxation time $T_1^{\rm Q}$; \cref{fig:figure4}(a) shows the result for all detected TLSs.
Since $\Gamma_{\rm relax}=\Gamma_1/2$ at resonance, subtracting the qubit contribution yields $T_1^{\rm TLS}=(2/T_{\rm relax}-1/T_1^{\rm Q})^{-1}$.
We estimated $T_1^{\rm Q}$ from single-exponential $T_1$ fits near the Rabi chevron.
In the example shown in~\cref{fig:figure3}, we found $T_1^{\rm Q}\approx380\ \mu$s on a qubit fabricated with atomic layer etching and deposition~\cite{Mahuli2025}, and $T_1^{\rm TLS}=16\pm3\ \mu$s.
Repeating this for every detected Rabi chevron, we characterized 118 TLSs, with $2g$ and $T_1^{\rm TLS}$ each spanning more than two orders of magnitude.
When $T_1^{\rm TLS}$ approaches or exceeds $T_1^{\rm Q}$, the subtraction introduces a large uncertainty in the estimate, which accounts for some of the largest error bars in~\cref{fig:figure4}(a) (see~\cref{sec:supp_fitting}).

The $T_1^{\rm TLS}$ values reported here are obtained in a coupling regime different from that of previous studies of coherent TLSs, and in a single FFA-SWAP spectroscopy sweep in which the qubit and TLS interact throughout the delay.
Coherent TLSs have previously been characterized with iSWAP-based protocols~\cite{Neeley2008,Shalibo2010,Chen2024}, in which a pair of calibrated swap pulses of tens of nanoseconds writes an excitation into the TLS and retrieves it after a variable delay, leaving it to relax in isolation.
Such protocols become impractical for the sub-MHz coupling rates studied here, where the swap pulse must last microseconds, a duration over which residual flux-pulse errors matter~\cite{Hellings2025}.
As shown in~\cref{fig:figure4}(a), our defects couple an order of magnitude more weakly than the JJ-barrier TLSs of previous work~\cite{Shalibo2010,Chen2024} (shaded regions).
A resolvable oscillation requires $2g$ to exceed $\Gamma_1^{\rm TLS}$, so this regime selects longer-lived defects, with a median $T_1^{\rm TLS}$ of approximately $10\ \mu$s.
Similar weakly coupled TLSs have been observed as incoherent $T_1$ dropouts, which yield the total decoherence rate and $g$ from a Lorentzian fit rather than from the swap-oscillation frequency~\cite{Klimov2018,Lisenfeld2019,Osman2023,Weeden2025}.

The rate $\Gamma_v=\Gamma_{\rm osc}-\Gamma_{\rm relax}$ extracted from each Rabi chevron further tests the dressed-state model.
We compared the measured $\Gamma_v(\delta\omega=0)$, resolved for 88 of the 118 TLSs, with the predicted qubit $1/T_2^*$; see~\cref{fig:figure4}(b).
We estimated $T_2^*$ assuming a $1/f$ flux-noise power spectrum~\cite{Anton2012,Yan2012}, with the flux susceptibility $\partial f_{01}/\partial\Phi$ evaluated at each TLS frequency.
We anchored the noise amplitude to the $T_2^*\approx1\ \mu$s measured in independent Ramsey experiments at $\partial f_{01}/\partial\Phi\approx5$~GHz$/\Phi_0$, close to the median $3.9$~GHz$/\Phi_0$ of the 88 TLSs (see~\cref{sec:supp_forward}).
Their medians are $0.2$ and $0.8\ \mu$s$^{-1}$, and $\Gamma_v<1/T_2^*$ for 77 of the 88 TLSs (dashed line).
This is evidence that the dressed gap suppresses the linear coupling to low-frequency flux noise~\cite{Makhlin2003,Muller2009}; simulations of the residual quadratic coupling account for the measured $\Gamma_v$ on about half of the cases, with the amplitude anchored instead to $T_2^*=0.5\ \mu$s (see~\cref{sec:supp_forward}).
Away from the TLS resonance the hybridization weakens, the linear coupling switches back on, and the measured $\Gamma_{\rm osc}(\delta\omega)$ rises toward the pure-dephasing rate $\Gamma_\phi$ of qubit and TLS.
The detuning dependence of $\Gamma_{\rm relax}$ further estimates the Markovian rate $(\Gamma_\pm+\Gamma_\mp)/2$, which could be subtracted from $\Gamma_v$ to isolate the non-Markovian part (see~\cref{sec:supp_dephasing}).

In conclusion, we have demonstrated that high-sensitivity SWAP spectroscopy enables the detection of Rabi chevrons at qubit-TLS coupling rates typically associated with the incoherent regime.
While the coherent/incoherent distinction is set by the coupling strength relative to the TLS relaxation rate~\cite{Barends2013}, observing coherent behavior at weak coupling also requires low dephasing and fine spectroscopic resolution.
We extend coherent detection to swap rates of $\sim0.1$~MHz, an order of magnitude lower than previously reported; in this band, the median TLS energy-relaxation time is $10\ \mu$s.
The long swap coherence times $T_{\rm osc}$ can be understood within the dressed-state picture, in which the dressed gap depends only quadratically on the qubit-frequency fluctuations produced by low-frequency noise that limits the qubit $T_2^*$.
This mechanism is analogous to that in spin-locked driven qubits~\cite{Ithier2005,Bylander2011,Yan2013}, two-qubit gates~\cite{Wang2026}, and transmon dual-rail architectures~\cite{Levine2024}, where a dressed gap protects the coherence against low-frequency noise.
This interpretation provides an alternative to reports that attribute long-lived coherent qubit-TLS oscillations to TLSs near their energy minima~\cite{Lisenfeld2016,Lisenfeld2019}.

The tools introduced here strengthen the growing effort in TLS engineering, from phononic bandgap structures~\cite{Chen2024} and strain- and electric-field tuning of TLSs~\cite{Lisenfeld2019} to investigations of possible links between charge TLSs and spin defects~\cite{Ateshian2025}.
An intriguing possibility is that TLSs classified as incoherent could be made to cross into the coherent regime.
Strain tuning can modify the TLS relaxation and dephasing rates~\cite{Lisenfeld2016,Lisenfeld2019}, and biasing the qubit near its flux-insensitive sweet spot reduces the dephasing noise that obscures slow chevron oscillations.
These considerations, together with previous limitations in the frequency and time resolution of SWAP spectroscopy~\cite{Barends2013,Klimov2018,Lisenfeld2019,Osman2023,Weeden2025,Lisenfeld2025}, will become important as TLS characterization scales to large device volumes~\cite{VanDamme2024,Ke2025,Mayer2025}.

\textit{Note added.}---During the preparation of this manuscript, a related work appeared that analyzes qubit-TLS swap oscillations with a dressed-state model including Markovian and non-Markovian noise~\cite{Gao2026}.

\begin{acknowledgments}
We thank the staff from across the Amazon Center for Quantum Computing that enabled this project.
A patent application related to this work has been filed.
\end{acknowledgments}

\bibliography{references}

\clearpage
\onecolumngrid
\begin{center}
  \large\bfseries Supplemental Material
\end{center}
\medskip
\twocolumngrid

\setcounter{secnumdepth}{3}
\setcounter{section}{0}
\setcounter{figure}{0}
\setcounter{equation}{0}
\setcounter{table}{0}
\renewcommand{\thesection}{S\arabic{section}}
\renewcommand{\thefigure}{S\arabic{figure}}
\renewcommand{\theequation}{S\arabic{equation}}
\renewcommand{\thetable}{S\arabic{table}}
\renewcommand{\theHfigure}{S\arabic{figure}}

\section{Chevron fitting and TLS parameter extraction}
\label{sec:supp_fitting}

For each frequency slice across a Rabi chevron, we fit the trace $P_1(\tau)$ by non-linear least squares, weighted by the per-point error bars, to the model function,
\begin{equation}
  P_1(\tau) = c + a_{\rm osc}\,\sin(\omega_{\rm osc}\tau+\phi)e^{-\tau/T_{\rm osc}}
              + a_{\rm relax}\,e^{-\tau/T_{\rm relax}},
  \label{eq:supp_model}
\end{equation}
as in~\cref{fig:figure3}(a).
We seed $\omega_{\rm osc}$ and $\phi$ from the strongest peak of a Lomb--Scargle (LS) periodogram~\cite{VanderPlas2018} of the same trace, or manually in the cases where that peak sits at very low frequency.
In those cases, the peak comes from the logarithmic delay spacing, which samples the long delays with few points so a slow sinusoid can carry large power, compounded by the relaxation term of~\cref{eq:supp_model}, since the periodogram removes only the mean of the trace and not the decay.
$T_{\rm osc}$ is seeded with the swap period $2\pi/\omega_{\rm osc}$ and $T_{\rm relax}$ with a single-exponential fit to the delays beyond the first oscillation cycles.
Both are bounded to the measured delay range, $0.1$--$10^{3}\ \mu$s or $0.1$--$10^{2}\ \mu$s depending on the qubit, and the amplitudes $a_{\rm osc}$ and $a_{\rm relax}$ are left free and non-negative.
A fit is accepted only if every parameter has a finite standard error, none reaches its bounds, the standard error is below the fitted value on $\omega_{\rm osc}$, $T_{\rm osc}$, $T_{\rm relax}$, $a_{\rm osc}$ and $a_{\rm relax}$, and the fitted $P_1(\tau)$ stays within $[0,1]$ over the full delay range.

\begin{figure}[t]
  \centering
  \includegraphics[width=\columnwidth]{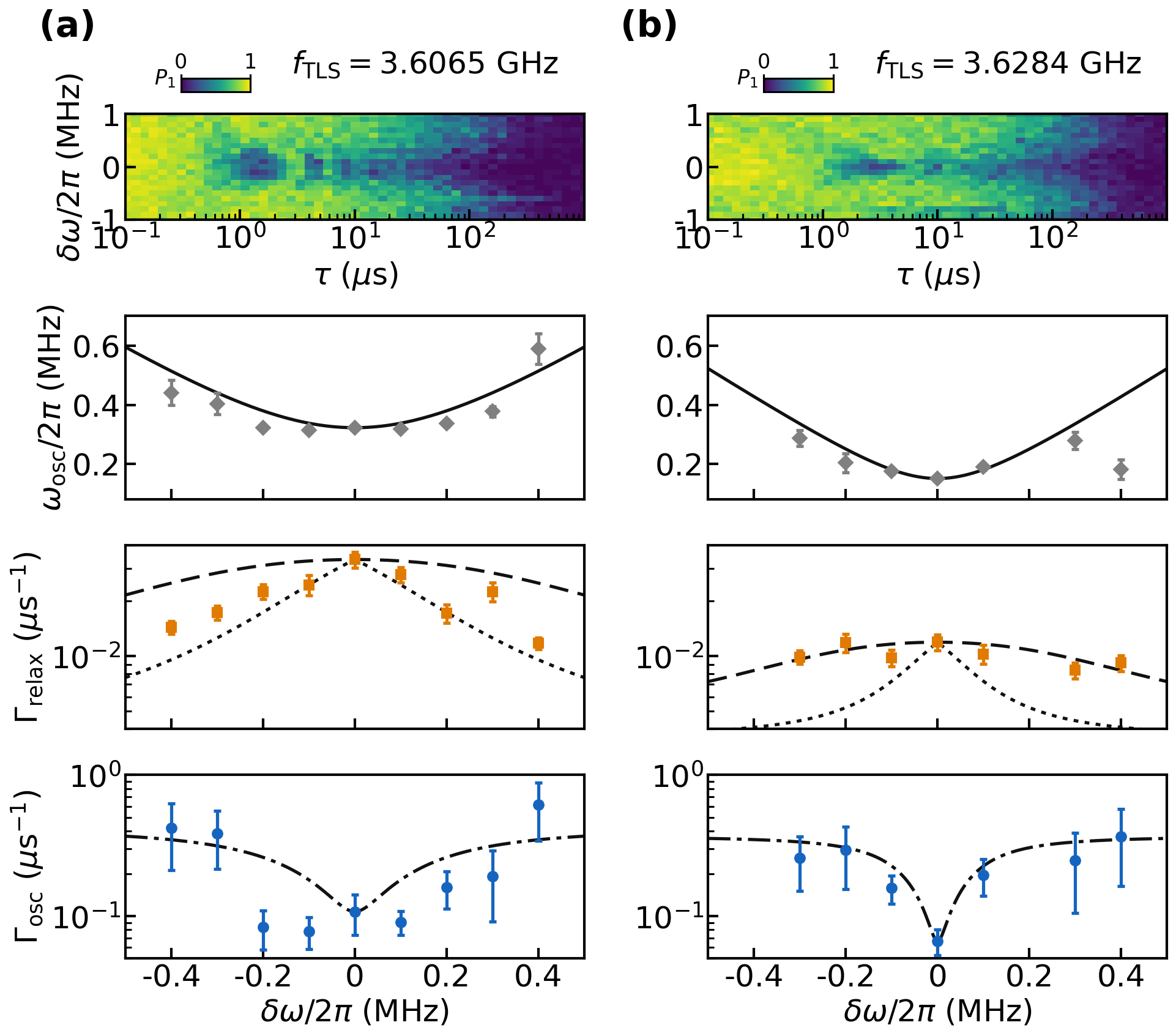}
  \caption{\protect\input{figureS1_caption.tex}}
  \label{fig:figureS1}
\end{figure}

We then plot $\omega_{\rm osc}$, $\Gamma_{\rm relax}$, and $\Gamma_{\rm osc}$ against the detuning $\delta\omega$ to locate the TLS resonance and extract the coupling $2g$, see~\cref{fig:figureS1}.
The oscillation frequency obeys the dispersion relation
\begin{equation}
  \omega_{\rm osc} = \sqrt{(2g)^2 + \delta\omega^2}.
  \label{eq:supp_dispersion}
\end{equation}
Selecting a slice as the resonance sets $\delta\omega=0$ and fixes $2g=\omega_{\rm osc}$; \cref{eq:supp_dispersion} is then drawn through the data (solid line, second row of~\cref{fig:figureS1}).
Where the minimum of $\omega_{\rm osc}$ alone is unclear, the two decay rates disambiguate the assignment: $\Gamma_{\rm relax}$ is maximal at resonance and falls on either side (third row), while we expect $\Gamma_{\rm osc}$ to reach its minimum there and rise on the flanks (bottom row), since pure dephasing of the swap oscillation switches off at resonance (see~\cref{sec:supp_dephasing}).
The two Rabi chevrons in~\cref{fig:figureS1} were detected on the same qubit at $f_{\rm TLS}=3.6065$ and $3.6284$~GHz, with couplings $2g/(2\pi)=0.32$ and $0.15$~MHz.
At $3.6065$~GHz [\cref{fig:figureS1}(a)] four adjacent slices share the same $\omega_{\rm osc}$ within their fit errors, and $\Gamma_{\rm relax}$, maximal at the selected slice, assigns the resonance.
At $3.6284$~GHz [\cref{fig:figureS1}(b)] $\Gamma_{\rm osc}$ assigns it instead, its minimum at the selected slice lying more than a factor of two below the adjacent ones.

\begin{figure}[t]
  \centering
  \includegraphics[width=0.95\columnwidth]{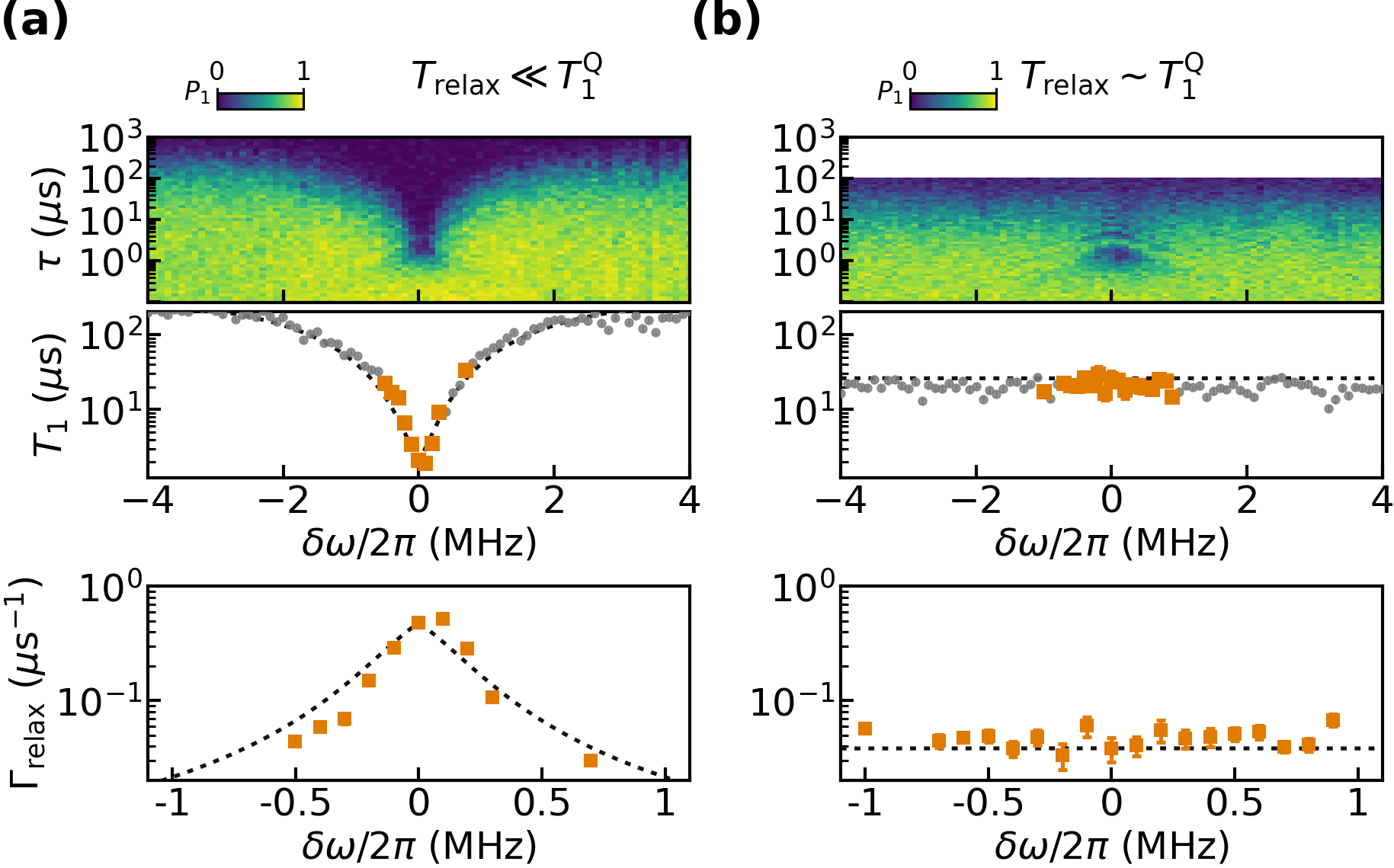}
  \caption{\protect\input{figureS2_caption.tex}}
  \label{fig:figureS2}
\end{figure}

In~\cref{fig:figure4}(a) we report the TLS energy-relaxation time $T_1^{\rm TLS}$ at the selected resonance slice, without interpolating between slices.
We use the relation $T_1^{\rm TLS}=(2/T_{\rm relax}-1/T_1^{\rm Q})^{-1}$, where the local $T_1^{\rm Q}$ is estimated per chevron from single-exponential fits to the slices flanking the resonance, excluding the dropout.
The error bar $\sigma(T_1^{\rm TLS})$ propagates the fit error on $T_{\rm relax}$.
The two Rabi chevrons shown in~\cref{fig:figureS2}, measured on two different qubits, illustrate the cases $T_{\rm relax}\ll T_1^{\rm Q}$ and $T_{\rm relax}\sim T_1^{\rm Q}$.
When $T_{\rm relax}\ll T_1^{\rm Q}$ [\cref{fig:figureS2}(a)], we find $T_{\rm osc}=T_{\rm relax}=2.1\ \mu$s, from the single-decay fit described below, and $T_1^{\rm TLS}=1.0\pm0.3\ \mu$s, far below $T_1^{\rm Q}=356\ \mu$s.
Notably, this is also a case where the swap oscillation does not complete a full cycle before it decays, since the swap period is $3.4\ \mu$s so $T_{\rm osc}<2\pi/(2g)$.
When $T_{\rm relax}\sim T_1^{\rm Q}$ [\cref{fig:figureS2}(b)], we find $T_1^{\rm TLS}=24\pm11\ \mu$s, where the large error bar follows from differencing two comparable rates.

The 118 Rabi chevrons of~\cref{fig:figure4} were detected on 16 qubits across 5 chips, grouped in~\cref{tab:devices} into four device groups, device~2 pooling two nominally identical chips.
The local $T_1^{\rm Q}$ is plotted per chevron in the top row of~\cref{fig:figureS3}(a) and varies from qubit to qubit even within one device group.
On device~1 it reaches $\approx380\ \mu$s on the qubit discussed in~\cref{fig:figure1,fig:figure2,fig:figure3}, and falls as low as $6\ \mu$s on the other two qubits, which sit closer to their readout resonators and are Purcell limited.
The accuracy of the subtraction therefore varies, and the 13 chevrons with $T_1^{\rm TLS}>T_1^{\rm Q}/2$ carry the largest error bars in~\cref{fig:figure4}(a).

In~\cref{fig:figure4}(b), we report the on-resonance rate $\Gamma_v=\Gamma_{\rm osc}-\Gamma_{\rm relax}$, with its error bar propagating the fit errors on $T_{\rm osc}$ and $T_{\rm relax}$.
The oscillatory component must decay at least as fast as the relaxation floor, so $\Gamma_v\ge0$, but the two decay times enter the fit independently, so a free fit can violate this bound or return a $\Gamma_v$ below its own uncertainty.
For 30 of the 118 cases we found $\Gamma_v/\sigma_{\rm fit}(\Gamma_v)<1$ and refit them with a single decay time, $P_1(\tau)=c+[a_{\rm osc}\sin(\omega_{\rm osc}\tau+\phi)+a_{\rm relax}]\,e^{-\tau/T_{\rm relax}}$, which sets $\Gamma_v=0$ exactly.
This has one parameter fewer than~\cref{eq:supp_model} and is favored by the Bayesian information criterion, in some cases by the full $\ln N\approx3.9$ expected from removing one parameter ($N\approx51$ delay points).
The remaining 88 keep their free fits and are the ones in~\cref{fig:figure4}(b).

\begin{table}[t]
  \squeezetable
  \caption{Device groups contributing the 118 Rabi chevrons of~\cref{fig:figure4}, in the order used as columns in~\cref{fig:figureS3,fig:figureS4}. The TLS column counts the chevrons in~\cref{fig:figure4}(a) and the next one those with a resolved $\Gamma_v$, which are the ones in~\cref{fig:figure4}(b) and in~\cref{fig:figureS3,fig:figureS4}.}
  \label{tab:devices}
  \begin{ruledtabular}
  \begin{tabular}{ccccc}
    Device & Process & Qubits & TLS & $\Gamma_v$ resolved \\
    \hline
    1 & planar ALE/ALD~\cite{Mahuli2025} & 3 & 35 & 25 \\
    2 & flip-chip                        & 7 & 26 & 18 \\
    3 & planar                           & 3 & 15 & 14 \\
    4 & planar                           & 3 & 42 & 31 \\
  \end{tabular}
  \end{ruledtabular}
\end{table}

\section{Dressed-state model of the swap decoherence}
\label{sec:supp_dephasing}

The dressed-state model of Ref.~\cite{Muller2009} describes the swap dynamics as a purely decaying part plus a damped oscillation.
Both respond to energy relaxation and to Markovian transitions between the dressed states, but only the oscillation is sensitive to pure dephasing from non-Markovian noise, so the detuning dependence of $\Gamma_{\rm relax}$ and $\Gamma_{\rm osc}$ separates the two channels.
The detuning enters both through the hybridization of the dressed states, set by the mixing angle $\xi$, with $\cos\xi=\delta\omega/\omega_{\rm osc}$ and $\sin\xi=2g/\omega_{\rm osc}$ (see~\cref{eq:supp_dispersion}).
At general detuning the purely decaying part contains two terms $a\,e^{-\Gamma_at}+b\,e^{-\Gamma_bt}$, with rates

\begin{equation}
  \begin{split}
    \Gamma_{a,b}=\frac{1}{2}\bigg[\,&\Gamma_1+2\Gamma_v^{\rm M}\sin^{2}\!\xi\\
      &\mp\sqrt{(\Gamma_1^{\rm TLS}-\Gamma_1^{\rm Q})^{2}\cos^{2}\!\xi
        +\left(2\Gamma_v^{\rm M}\sin^{2}\!\xi\right)^{2}}\;\bigg],
  \end{split}
  \label{eq:gamma_relax}
\end{equation}
where $\Gamma_1=\Gamma_1^{\rm Q}+\Gamma_1^{\rm TLS}$ is the total energy-relaxation rate, $\Gamma_v^{\rm M}\sin^{2}\!\xi\equiv\tfrac12(\Gamma_\pm+\Gamma_\mp)$, and $\Gamma_\pm$, $\Gamma_\mp$ are the transition rates between the dressed states driven by Markovian noise at the gap~\cite{Muller2009} (see~\cref{fig:figure3}(b)).
We take $\Gamma_\pm=\Gamma_\mp$: the dressed gap $\omega_{\rm osc}/2\pi\lesssim22$~MHz lies far below $k_BT/h\approx400$~MHz at $T=20$~mK, so the noise at the gap is effectively classical.

Our fit function~\eqref{eq:supp_model} uses a single purely decaying term, so the extracted $\Gamma_{\rm relax}$ is an effective rate.
On resonance ($\delta\omega=0$) the amplitude $b$ vanishes and $\Gamma_a=\Gamma_1/2$ exactly, so $\Gamma_{\rm relax}=\Gamma_1/2$ independently of $\Gamma_v^{\rm M}$, the relation used to extract $T_1^{\rm TLS}$ in~\cref{fig:figure4}(a).
Far from the TLS resonance, $\vert\delta\omega\vert\gg2g$, $\sin\xi\to0$, so $\Gamma_v^{\rm M}$ drops out and $\Gamma_{a,b}$ reduce to the bare rates $\Gamma_1^{\rm Q}$ and $\Gamma_1^{\rm TLS}$, with $\Gamma_a\to\Gamma_1^{\rm Q}$ if $\Gamma_1^{\rm TLS}\geq\Gamma_1^{\rm Q}$.
In the $\Gamma_{\rm relax}$ panels of~\cref{fig:figureS1} (orange squares), the dotted curves are the rate $\Gamma_a$ of~\eqref{eq:gamma_relax} at $\Gamma_v^{\rm M}=0$.
They are evaluated, not fitted, with $2g$ and $\Gamma_1$ from the TLS resonance and $\Gamma_1^{\rm Q}$ from the single-exponential fits flanking the chevron.
For the two chevrons of~\cref{fig:figureS2}, the dotted lines ($\Gamma_v^{\rm M}=0$) in the $T_1$ (middle row) and $\Gamma_{\rm relax}$ (bottom row) panels follow the measured data, consistent with a negligible Markovian contribution $\Gamma_v^{\rm M}$ at both resonances.
For $\vert\delta\omega\vert\gg2g$ they also agree with the qubit $T_1$ from single-exponential fits (grey circles).
In both chevrons of~\cref{fig:figureS1}, however, the dotted curves underestimate the measured $\Gamma_{\rm relax}$, which could be evidence for a finite $\Gamma_v^{\rm M}$.

Markovian noise also enters the swap-oscillation decay as an exponential law $e^{-(\Gamma_1/2+\Gamma_v^{\rm M}\sin^2\!\xi)t}$, and linear coupling to non-Markovian $1/f$ noise enters as a Gaussian decay, $e^{-\Gamma_\phi^2\cos^{2}\!\xi t^2}$, with $\Gamma_\phi$ the pure-dephasing rate.
The two combine in the envelope,
\begin{equation}
  -\ln|f(t)|=\Big(\frac{\Gamma_1}{2}+\Gamma_v^{\rm M}\sin^{2}\!\xi\Big)t
    +\Gamma_\phi^{2}\cos^{2}\!\xi t^{2},
  \label{eq:envelope}
\end{equation}
whose $1/e$ point defines the measured rate through $-\ln|f(1/\Gamma_{\rm osc})|=1$.
At the TLS resonance $\cos\xi\to0$, so the pure-dephasing term switches off and $\Gamma_{\rm osc}\to\Gamma_1/2+\Gamma_v^{\rm M}$; with $\Gamma_{\rm relax}=\Gamma_1/2$, the model gives $\Gamma_v=\Gamma_v^{\rm M}$.
The dash-dotted curves in the bottom row of~\cref{fig:figureS1} are the prediction of~\cref{eq:envelope} with $\Gamma_v^{\rm M}=\Gamma_v$ and $\Gamma_\phi=1/T_2^{*}$, with the qubit $T_2^{*}$ typically limited by linear coupling to flux noise.
The $\Gamma_{\rm relax}$ panels of~\cref{fig:figureS1} carry a second, dashed curve: \cref{eq:gamma_relax} evaluated at the upper bound $\Gamma_v^{\rm M}=\Gamma_v$.
For the chevron at $f_{\rm TLS}=3.6284$~GHz [\cref{fig:figureS1}(b)] the dashed line agrees with the measured $\Gamma_{\rm relax}$, so most of $\Gamma_v$ is Markovian, $\Gamma_v\approx\Gamma_v^{\rm M}$.
At $3.6065$~GHz [\cref{fig:figureS1}(a)], however, the dashed line overestimates the measured $\Gamma_{\rm relax}$ while the dotted line underestimates it.
This indicates an additional dephasing contribution at the TLS resonance, modeled below as quadratic coupling of the swap frequency to the qubit's $1/f$ flux noise~\cite{Muller2009,Makhlin2003}.

We model the qubit flux noise as a fluctuating term in the non-interacting qubit Hamiltonian, $H_q/\hbar=\tfrac{1}{2}[\omega_{01}+\delta\omega_{01}(t)]\sigma_z$, with $\delta\omega_{01}(t)=(\partial\omega_{01}/\partial\Phi)\,\delta\Phi(t)$, where $\partial\omega_{01}/\partial\Phi$ is the flux susceptibility and $\delta\Phi(t)$ the flux fluctuations.
For a $1/f$ spectrum, components slower than the inverse duration of a single shot carry most of the noise power and act as a quasi-static flux offset.
It is a stationary Gaussian process with two-sided spectral density $S_\Phi(f)=A_\Phi/|f|$ ($\Phi_0^2/{\rm Hz}$), normalized so that $\langle\delta\Phi^2\rangle=\int_{-\infty}^{\infty}S_\Phi(f)\,df$, where $A_\Phi$ is the noise power at $1\,$Hz; we quote the amplitude as $\sqrt{A_\Phi}$ in $\mu\Phi_0$ units~\cite{Ithier2005,Anton2012}.
For a detuning $\delta\omega$, the instantaneous swap frequency $\omega_{\rm osc}=\sqrt{(2g)^2+[\delta\omega+\delta\omega_{01}(t)]^2}$ expands to second order as
\begin{equation}
  \omega_{\rm osc}(t)\simeq\omega_0
    +\cos\xi\,\delta\omega_{01}(t)
    +\frac{\sin^2\!\xi}{2\,\omega_0}\,\delta\omega_{01}(t)^2,
  \label{eq:wosc_expand}
\end{equation}
with $\omega_0=\sqrt{(2g)^2+\delta\omega^2}$ and $\xi$ as above.
Averaging the accumulated phase $\phi(t)=\int_0^t\omega_{\rm osc}(t')\,dt'$ over flux realizations, the linear term in $\delta\omega_{01}(t)$ gives the Gaussian law $e^{-\Gamma_\phi^2\cos^{2}\!\xi\,t^2}$, with $\Gamma_\phi^2=(\partial\omega_{01}/\partial\Phi)^2A_\Phi\ln[1/(f_1t)]$, where $f_1$ is the infrared cut-off of the Ramsey decay.
The quadratic term $\delta\omega_{01}(t)^2$ gives, at long times, an exponential decay law $\exp(-\Gamma_f t/2)$ with $\Gamma_f/2=\tfrac14\sin^2\xi (\partial\omega_{01}/\partial\Phi)^2A_\Phi (2\pi/\omega_0)$~\cite{Makhlin2003}.
The quadratic coupling thus adds $\Gamma_ft/2$ to the right-hand side of~\eqref{eq:envelope}; a similar term would appear for low-frequency noise local to the TLS.
At the TLS resonance $\cos\xi=0$, $\sin\xi=1$ and $\omega_0=2g$, so the linear term switches off and only the quadratic one survives, giving the qubit's contribution to the dressed-state pure dephasing,
\begin{equation}
  \frac{\Gamma_f}{2}
    =\frac14\left(\frac{\partial\omega_{01}}{\partial\Phi}\right)^{\!2}
     A_\Phi\left(\frac{2\pi}{2g}\right),
  \label{eq:gammaf_quad}
\end{equation}
so the residual dephasing scales as $1/2g$: a larger gap protects the swap oscillation more strongly.

\Cref{eq:gammaf_quad} coincides symbolically with the Fermi-golden-rule (FGR) rate $\tfrac14(\partial\omega_{01}/\partial\Phi)^2S_\Phi[2g/(2\pi)]$, since extrapolating the low-frequency spectrum to the gap would give $S_\Phi[2g/(2\pi)]=A_\Phi/[2g/(2\pi)]$.
The coincidence is a feature of the $1/f$ shape and does not survive for the experimentally relevant $1/f^\alpha$ spectra with $\alpha<1$~\cite{Ateshian2025}; more importantly, the two rates sample different parts of the spectrum.
The FGR formula weights the noise at the gap frequency $2g/(2\pi)\gtrsim0.1$~MHz, accessible to spin-locking and CPMG sequences.
By contrast, \eqref{eq:gammaf_quad} follows entirely from the low-frequency $1/f$ band, which can be measured with repeated Ramsey experiments at fixed pulse separation~\cite{Yan2012}.

\begin{figure*}[t]
  \centering
  \includegraphics[width=\textwidth]{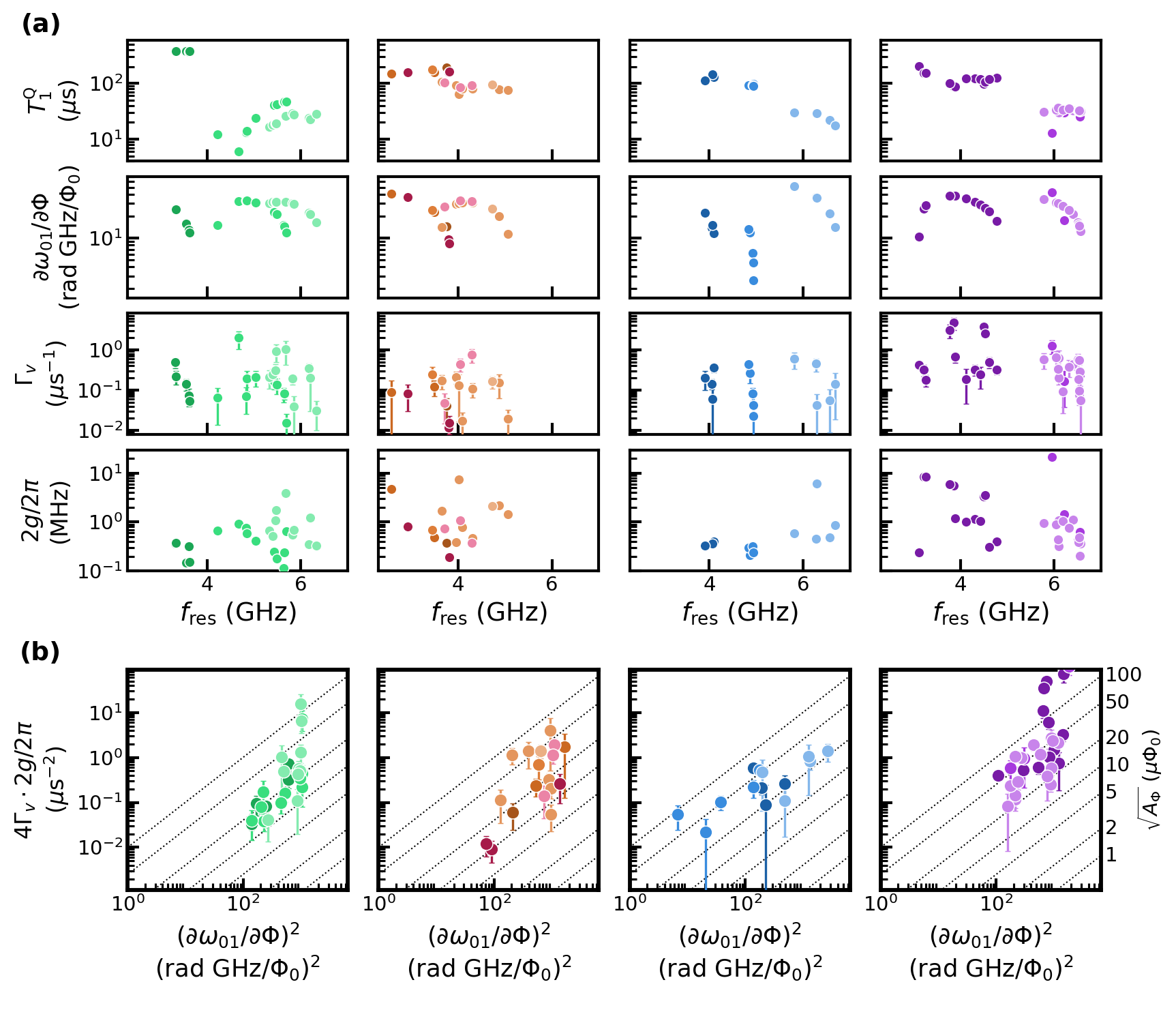}
  \caption{\protect\input{figureS3_caption.tex}}
  \label{fig:figureS3}
\end{figure*}

To test the predicted dephasing from the quadratic coupling, we rewrite~\eqref{eq:gammaf_quad} as $4(\Gamma_f/2)\cdot 2g/(2\pi)=(\partial\omega_{01}/\partial\Phi)^2A_\Phi$, which isolates the flux noise from the gap.
Were the qubit's flux noise the sole source of the measured $\Gamma_v$, plotting $4\Gamma_v\cdot 2g/(2\pi)$ against $(\partial\omega_{01}/\partial\Phi)^2$ would place every chevron on a diagonal of constant $A_\Phi$.
The diagonal through each point gives the $\sqrt{A_\Phi}$ at which the quadratic coupling alone reproduces its measured $\Gamma_v$.
\Cref{fig:figureS3} shows this one column per device group, devices~1 to~4 from left to right, so each group can be inspected separately.
The implied amplitudes from~\cref{fig:figureS3}(b) have a median $\sqrt{A_\Phi}=33\ \mu\Phi_0$, high compared with reported values~\cite{Anton2012,Ateshian2025}.
Since the measured $\Gamma_v$ also contains $\Gamma_v^{\rm M}$ and any dephasing from noise local to the TLS, the amplitude read off each diagonal is an upper bound on the qubit's flux noise.

\section{Dephasing prediction from an independently calibrated flux amplitude}
\label{sec:supp_forward}

\begin{figure*}[t]
  \centering
  \includegraphics[width=\textwidth]{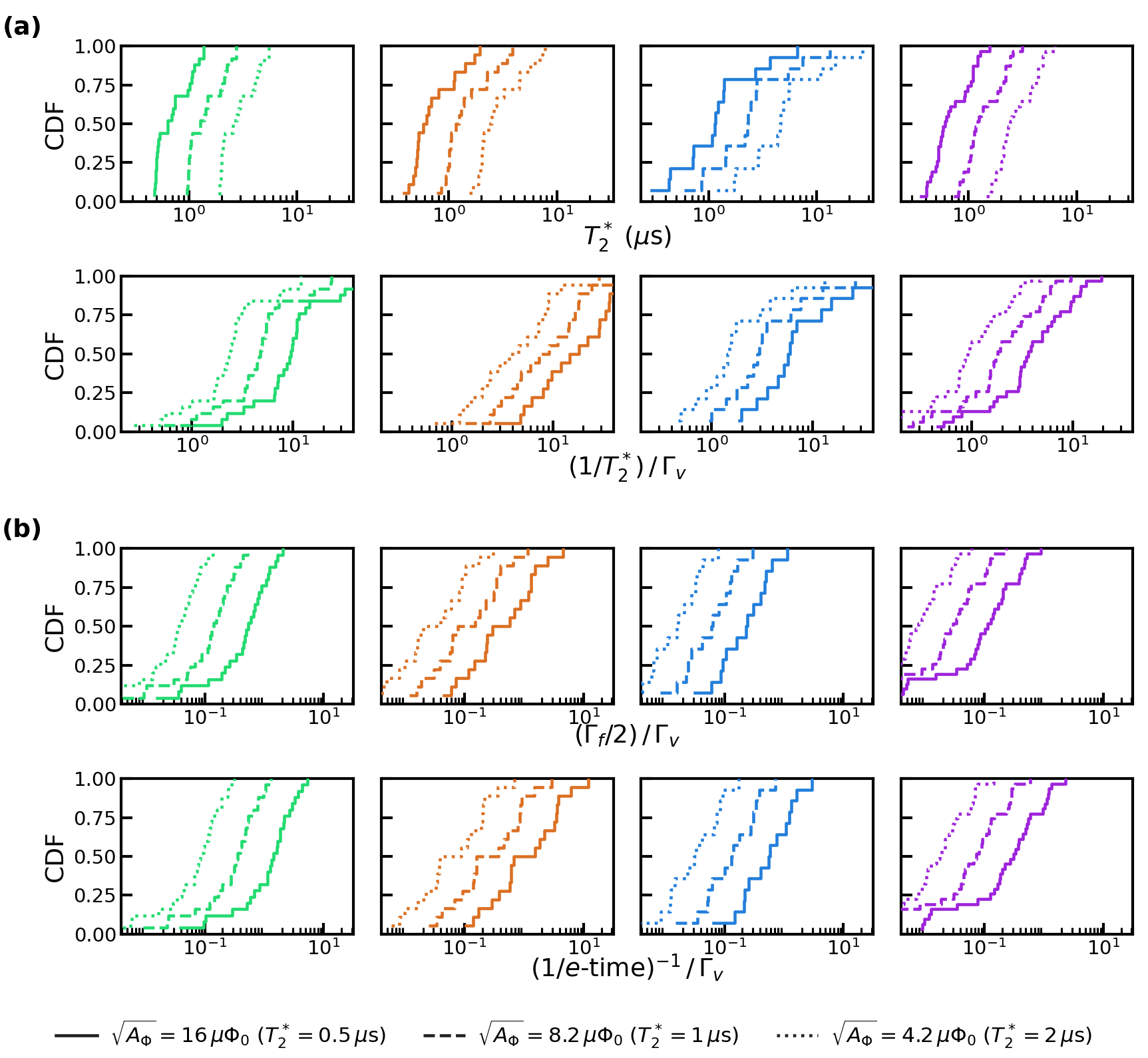}
  \caption{\protect\input{figureS4_caption.tex}}
  \label{fig:figureS4}
\end{figure*}

We now fix the flux-noise amplitude $A_\Phi$ at realistic experimental values and compare the dephasing predicted from the quadratic coupling with the $\Gamma_v$ measured at the 88 resolved Rabi chevrons.
The flux noise seen by each Rabi chevron can be calibrated from independent Ramsey measurements, in which qubit times $T_2^{*}\approx1\,\mu$s were found at flux susceptibilities $(1/2\pi)\,\partial\omega_{01}/\partial\Phi\approx5$~GHz$/\Phi_0$.
Because $T_2^*$ is not measured at every chevron, we anchor the amplitude to three values, $T_2^*\in\{2,1,0.5\}\,\mu$s.
For each anchor we solve $\exp(-\Gamma_\phi^2t^2)=1/e$ at $t=T_2^{*}$ for the amplitude $A_\Phi$ that reproduces it at the reference susceptibility $(1/2\pi)\,\partial\omega_{01}/\partial\Phi=5$~GHz$/\Phi_0$, with infrared cutoff $f_1=0.1$~Hz.
For the $T_2^{*}=2$, $1$ and $0.5\,\mu$s anchors this gives $\sqrt{A_\Phi}=4.2$, $8.2$ and $16\ \mu\Phi_0$ respectively, the amplitudes labeling the curves in~\cref{fig:figureS4}.
We then propagate that single amplitude through the susceptibility $\partial\omega_{01}/\partial\Phi$ expected at each Rabi chevron, integrating the flux-noise band up to $f_2=2g/(2\pi)$, the upper edge of the band in which $\delta\omega_{01}(t)$ enters only as a slow modulation of the swap frequency.
The calibration is confirmed in the upper row of~\cref{fig:figureS4}(a), one column per device group, devices~1 to~4 left to right and with traces for each anchor $T_2^{*}=2,1,0.5\ \mu$s (dotted, dashed, solid lines).
The $T_2^{*}$ values predicted for all Rabi chevrons cluster at the anchor, spread only by the range of $\partial\omega_{01}/\partial\Phi$; the tail towards long $T_2^{*}$ comes from the small-susceptibility chevrons, for which a fixed flux amplitude implies a longer Ramsey time.
The central result is in the lower row of~\cref{fig:figureS4}(a): the predicted qubit Ramsey rate typically exceeds the measured dressed-gap dephasing rate, with medians for $(1/T_2^{*})/\Gamma_v$ of 1.75, 3.50 and 7.01 for the $T_2^{*}=2$, $1$ and $0.5\,\mu$s anchors respectively.
For the nominal anchor $T_2^{*}=1\,\mu$s the two distributions have medians $\Gamma_v=0.2\ \mu$s$^{-1}$ and $1/T_2^{*}=0.8\ \mu$s$^{-1}$, the values plotted in~\cref{fig:figure4}(b), where $\Gamma_v<1/T_2^{*}$ for 77 of the 88 chevrons.

\cref{fig:figureS4}(b) shows the predicted dephasing rate relative to the measured rate, one column for each device group showing the cumulative distribution functions for the three $T_2^{*}$ anchors as in~\cref{fig:figureS4}(a), with two evaluations of the quadratic coupling to qubit flux noise.
The top row uses the \emph{long-time slope} $\Gamma_f/2$ of~\eqref{eq:gammaf_quad}; the bottom row uses the inverse of the $1/e$ time obtained from the Gaussian quadratic form for the truncated phase $\phi=\int 2g\ x^2(t')/2\ dt'$, with $x=[(\partial\omega_{01}/\partial\Phi)\delta\Phi]/(2g)$.
In the bottom row, with the amplitude anchored to $T_2^*=0.5\ \mu$s, the column medians of the predicted-to-measured ratio run $1.45$, $1.09$, $0.57$ and $0.31$ across devices~1 to~4, a monotone device gradient: the prediction reaches the measured $\Gamma_v$ on devices~1 and~2 and falls a factor of two to three short on devices~3 and~4.
The two methods differ because the long-time slope is an asymptotic result, whereas the actual decay can reach its $1/e$ point before that asymptotic regime sets in~\cite{Makhlin2003}.
The truncation $x^2/2$ also overestimates the exact excursion $\sqrt{1+x^2}-1$, so the truncated phase overestimates the dephasing and is an upper bound on the prediction.
In the future, Monte Carlo simulations of the untruncated phase, and a repeat of the analysis for $1/f^\alpha$ spectra with $\alpha<1$, would give more realistic estimates.
Also, a measured qubit noise power spectrum would fix the linear-coupling rate $\Gamma_\phi$ independently, so that the dependence of $\Gamma_{\rm osc}$ on $\delta\omega$ could isolate the TLS contribution to the dephasing, and it would constrain the Markovian part $\Gamma_v^{\rm M}$ separately, leaving a cleaner test of the quadratic coupling term.

\end{document}